\documentclass[9pt]{article}
\usepackage{natbib}
\usepackage{graphicx}
\usepackage{amsmath,amssymb}
\usepackage{xcolor}
\usepackage{float}
\usepackage{array}
\usepackage{adjustbox}
\usepackage{booktabs}
\usepackage{lineno}
\usepackage{algorithm}
\usepackage{algpseudocode}
\usepackage{subcaption}
\usepackage{microtype}
\usepackage{makecell}
\usepackage{tabularx}
\usepackage{url}

\title{Professional networks and the diffusion of clinical guidelines in opioid prescribing}

\author{
Yi-Ning Weng \\
Department of Accounting, National Taiwan University \\
\and
Hsuan-Wei Lee \\
College of Health, Lehigh University \\
\texttt{hsl324@lehigh.edu}
}

\date{}

\begin{document}

\maketitle

\begin{abstract}

Large and persistent differences in opioid prescribing across physicians and regions cannot be explained by patient characteristics or physician attributes alone. We developed a behavioral framework in which prescribing evolves through persistence, exposure to peers in professional networks, and heterogeneous responses to a common policy signal that varies with network centrality. Using nationwide Medicare Part D data from 2013 to 2020, covering more than two million physician-year observations, we tested three hypotheses implied by this framework. Physicians exposed to higher peer prescribing subsequently prescribe more; more central physicians reduce prescribing more following the introduction of the 2016 CDC guideline, with no evidence of differential pre-trends; and changes in peer prescribing are closely associated with changes in individual prescribing in the post-guideline period. By 2020, physicians at the 90th percentile of network centrality exhibited prescribing reductions 0.30 percentage points larger than those at the 10th percentile, with the gap widening steadily after the introduction of the CDC guideline. Together, these results indicate that opioid prescribing operates through professional networks, in which policy effects spread through connections and appear to be shaped by network position. This suggests that engaging highly connected physicians may help extend the reach of opioid stewardship programs. It also raises questions about how the burden and benefits of such targeting would be distributed across physicians and patients.

\end{abstract}


\section{Introduction}

The opioid epidemic has claimed more than 500,000 lives in the United
States over the past two decades, and physician prescribing behavior has
played a central role in its trajectory. Despite extensive research, large
and persistent differences in prescribing intensity across physicians and
regions remain difficult to explain, even after accounting for patient
characteristics and the clinical context. Between 2007 and 2012, the share
of Medicare Part D recipients receiving prolonged opioid therapy nearly
doubled, and substantial geographic variation persisted after adjusting for
patient factors \citep{kuo2016trends}. Standard explanations emphasize the
patient case mix, regulatory environments, and physician training; however,
these factors do not fully account for the observed heterogeneity. A
central question remains: why do physicians serving similar patients in
similar settings prescribe so differently?

A growing body of literature suggests that professional networks may
explain this variation. Physicians operate in environments where peer
behavior, shared norms, and local information flow shape clinical decisions
\citep{tasselli2014social,cunningham2012health}. Evidence from
pharmaceutical adoption indicates that physicians are more likely to
prescribe new drugs when their peers have already done so, particularly
when they share patients or practice in close proximity
\citep{donohue2018influence,nair2010asymmetric,yang2014there}. More
centrally positioned physicians often act as opinion leaders and influence
the diffusion of clinical practice
\citep{soumerai1998effect,locock2001understanding,hao2018empirical}.
Similar patterns have been documented in antibiotic prescribing
\citep{gong2025peer,wang2023can}. Whether these network mechanisms operate
in opioid prescribing and shape responses to major policy interventions
remains unclear. In particular, two questions remain unresolved. Do peer
effects persist after accounting for physician fixed effects and local
shocks? Does network position influence how physicians respond to national
clinical guidelines?

This study addresses these questions by developing a unified behavioral
framework that links persistence in individual prescribing, exposure to
peer behavior, and heterogeneous responses to a common policy signal. This
framework builds on the DeGroot tradition of network-based social learning
\citep{golub2010naive,chandrasekhar2020testing,choi2023learning}. A key
feature is that the responsiveness to external policy signals varies with
network centrality, reflecting differences in exposure to peers and
information flows \citep{kandhway2016using}. The network structure plays
two roles: determining the extent of peer exposure and shaping how policy
signals are translated into behavioral changes.

The framework yields three testable hypotheses. First, physicians exposed
to higher peer prescribing intensity exhibit higher subsequent prescribing,
reflecting social learning within local professional networks
\citep{gong2025peer,wang2023can,yang2014there}. Second, physicians who are
more centrally positioned in the network respond more strongly to common
policy shocks because their greater exposure to peer behavior amplifies the
transmission of external signals. Third, changes in peer prescribing are
associated with changes in individual prescribing following policy
intervention, consistent with network-mediated propagation of behavioral
adjustment \citep{kreindler2014diffusion,onnela2010influence}.

These hypotheses were evaluated using nationwide Medicare Part D
prescribing data from 2013 to 2020
\citep{donohue2012sources,kuo2016trends,yin2008effect,ang2025expanded}.
The 2016 CDC opioid prescribing guideline provides a well-defined national
policy shock \citep{bohnert2018opioid}. Physician networks are constructed
from geographic co-location within ZIP codes, a tractable proxy for local
professional interaction
\citep{tasselli2015social,cunningham2012health}.

Our empirical analysis proceeded in three steps. First, we documented a
positive association between peer and individual prescribing robust across
specifications with physician and county-by-year fixed effects. Second,
using an event study design around the 2016 CDC guideline, we showed that
physicians with higher network centrality exhibited larger reductions in
prescribing following the policy change, with no evidence of differential
pre-trends. Third, we found that changes in peer prescribing were
positively associated with changes in individual prescribing in the
immediate post-guideline period.

These results are consistent with a process in which peer exposure, network position, and policy responses together shape prescribing behavior. Policy signals do not act on physicians in isolation,
but move through professional connections, generating heterogeneous and
evolving responses across the network. Consequently, policies that account
for professional network structures may produce broader and more sustained
reductions in prescribing than interventions that target physicians
individually
\citep{valente2012network,galeotti2020targeting,cabana1999don,kilaru2014physicians}.

Beyond opioid prescribing, this study speaks to a broader question in the social science of medicine. It asks how professional norms and peer relationships shape the way clinicians respond to external policy. The physician setting lets us observe this social process at large scale and connect it to a major public health intervention.

\section{Methods}

\subsection*{Data and Sample Construction}

Our analysis drew on three publicly available data sources. Our primary
data source was the Medicare Part D Prescriber Public Use Files provided by
the Centers for Medicare and Medicaid Services, covering 2013 to 2020. The
unit of observation was physician-year.

We restricted the sample to seven commonly prescribed opioid analgesics,
which together accounted for approximately 71\% of all opioid claims
(Table~\ref{tab:opioids}). Drugs were selected based on four criteria:
active opioid ingredients classified under ICD-10 code T40.2, mappable
morphine milligram equivalent (MME) conversion factors, consistent
appearance across all sample years in the Medicare Part D data, and common
use as analgesics for routine outpatient pain management. All prescriptions
were converted to MME using conversion factors from the 2022 CDC Clinical
Practice Guideline for Prescribing Opioids for Pain. Physician geographic
information was obtained from prescriber ZIP codes in the CMS data, mapped
to counties using HUD--USPS ZIP Code Crosswalk files.

The primary outcome was prescribing intensity, defined as the total MME
prescribed divided by the number of opioid claims in a given physician-year.
Patient volume was measured as the total number of unique Medicare
beneficiaries associated with each physician in a given year.

\begin{table}[H]
\centering
\caption{Selected opioid drugs and their share of total opioid claims, 2013--2020.}
\label{tab:opioids}
\begin{tabular}{lrr}
\toprule
Drug name & Total claims & Share (\%) \\
\midrule
Hydrocodone/Acetaminophen     & 213,122,922 & 39.68 \\
Oxycodone HCl/Acetaminophen   &  69,842,984 & 13.00 \\
Oxycodone HCl                 &  60,481,680 & 11.26 \\
Morphine Sulfate              &  27,303,780 &  5.08 \\
Hydromorphone HCl             &   6,436,994 &  1.20 \\
Oxymorphone HCl               &   1,964,177 &  0.37 \\
Hydrocodone/Ibuprofen         &     672,224 &  0.13 \\
\midrule
\textbf{Total (7 drugs)}      & \textbf{379,824,761} & \textbf{70.72} \\
\bottomrule
\end{tabular}
\end{table}

\subsection*{Network Construction and Centrality}

Annual physician networks were constructed based on geographic proximity.
Two physicians were connected in a given year if they practiced within the
same ZIP code. The weight between physicians $i$ and $j$ was:
\begin{equation}
w_{ij} = 0.1 \times \min(\text{Tot\_Benes}_i, \text{Tot\_Benes}_j),
\end{equation}
where $\text{Tot\_Benes}_i$ denotes the number of unique Medicare
beneficiaries treated by physician $i$. Network position was measured using
weighted degree centrality, defined as the sum of a physician's edge
weights.

\subsection*{Unified Behavioral Framework}

We modeled physician prescribing as a dynamic updating process following
the DeGroot tradition
\citep{golub2010naive,degroot1974consensus,ding2019consensus}:
\begin{equation}
p_{i,t+1} = \alpha p_{i,t} + \lambda \bar{p}_{N(i),t} + \mu_i \theta G_t + \varepsilon_{i,t+1}.
\end{equation}
Here $p_{i,t}$ denotes prescribing intensity, $\bar{p}_{N(i),t}$ is the
weighted average prescribing intensity of physician $i$'s network peers,
$\alpha$ captures persistence, $\lambda$ captures social learning, and
$G_t$ denotes the external signal corresponding to the 2016 CDC guideline.
We assumed $\alpha + \lambda < 1$. We allowed centrality-dependent
responsiveness:
\begin{equation}
\mu_i = \mu_0 + \mu_1 C_i,
\end{equation}
where $C_i$ denotes physician $i$'s network centrality. We held $\lambda$
constant across physicians and focused inference on $\mu_1$.

\subsection*{Testable Hypotheses and Empirical Specifications}

\textit{Hypothesis~1.} Physicians exposed to higher peer prescribing
subsequently prescribe more:
\begin{equation}
\log(1+p_{i,t+1}) = \beta_1 \log(1+\bar{p}_{N(i),t}) + \beta_2 \log(1+p_{i,t}) + \alpha_i + \lambda_{\mathrm{county}(i),t} + X'_{i,t}\gamma + \varepsilon_{i,t}.
\end{equation}
A positive $\beta_1 > 0$ supports this hypothesis.

\textit{Hypothesis~2.} Physicians with higher network centrality exhibit
greater reductions in prescribing following the policy intervention:
\begin{equation}
\log(1+p_{i,t}) = \sum_{\tau \neq 2015} \delta_\tau \bigl(1[t=\tau]\times C_i\bigr) + \alpha_i + \lambda_t + X'_{i,t}\gamma + \varepsilon_{i,t}.
\end{equation}
$C_i$ is measured using the 2015 network and standardized as a z-score.
Log-transformed centrality violates the parallel trends assumption
($p = 0.0002$) and is not adopted. More negative post-2016 $\delta_\tau$
coefficients are consistent with this hypothesis.

\textit{Hypothesis~3.} Physicians whose peers reduce prescribing more
exhibit larger own reductions:
\begin{equation}
\Delta p_i = \beta\,\Delta \bar{p}_{N(i)} + \Gamma'X_{i,2016} + \varepsilon_i,
\end{equation}
where $\Delta p_i$ and $\Delta \bar{p}_{N(i)}$ denote changes in
individual and peer prescribing between 2016 and 2017. The estimation
sample consists of 246,481 physicians observed in both years with
non-missing peer prescribing changes. Across all specifications, standard
errors are clustered at the county level.

\section{Results}

\subsection*{Network Structure and Aggregate Prescribing Patterns}

Table~\ref{tab:network_summary} reports summary statistics for the same-ZIP
physician network and opioid prescribing behavior from 2013 to 2020. The
network was large and densely connected throughout the sample period, with
more than 300,000 physicians per year and an average degree exceeding 80
even in later years. The network remained highly stable over time, with
node Jaccard similarity around 0.70--0.75 across adjacent years. Total MME
declined steadily after 2016, consistent with the CDC guideline timing.

\begin{table}[H]
\centering
\caption{Summary statistics of same-ZIP physician network structure and opioid prescribing behavior, 2013--2020.
Nodes correspond to physicians; edges indicate co-location within the same ZIP code.
Avg.\ degree is the average number of connections per physician.
Total MME claims measure aggregate opioid prescribing volume.
MME per prescriber is the average prescribing volume per physician.
MME per beneficiary is total MME-weighted claims divided by total beneficiaries.}
\label{tab:network_summary}
\begin{adjustbox}{width=\textwidth}
\begin{tabular}{lrrrrrr}
\toprule
Year & Nodes & Edges & Avg.\ degree & Total MME claims & MME per prescriber & MME per beneficiary \\
\midrule
2013 & 376,021 & 19,411,088 & 103.25 & 63,198,978 & 168.07 & 0.784 \\
2014 & 378,963 & 19,851,482 & 104.77 & 65,115,840 & 171.83 & 0.770 \\
2015 & 362,141 & 18,241,140 & 100.74 & 64,048,801 & 176.86 & 0.758 \\
2016 & 358,240 & 17,987,489 & 100.42 & 63,675,193 & 177.75 & 0.743 \\
2017 & 346,564 & 16,869,380 &  97.35 & 60,439,886 & 174.40 & 0.708 \\
2018 & 324,740 & 14,856,886 &  91.50 & 55,607,602 & 171.24 & 0.668 \\
2019 & 310,098 & 13,758,441 &  88.74 & 51,912,316 & 167.41 & 0.634 \\
2020 & 294,899 & 12,447,808 &  84.42 & 49,981,873 & 169.49 & 0.654 \\
\bottomrule
\end{tabular}
\end{adjustbox}
\end{table}

\begin{figure}[H]
\centering
\includegraphics[width=0.95\textwidth]{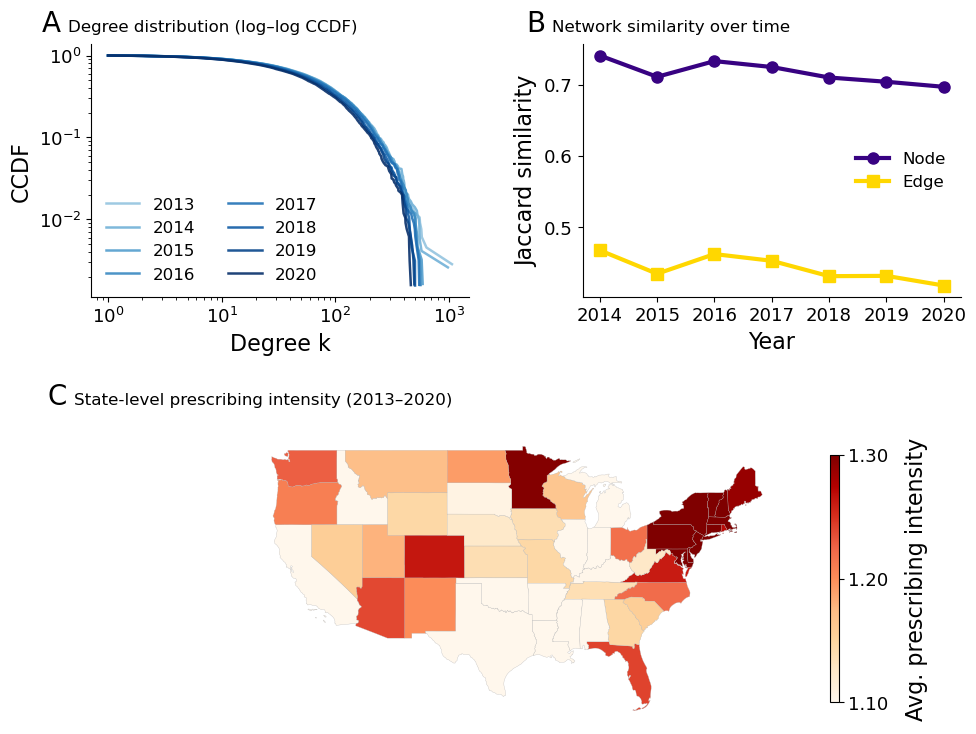}
\caption{Structure and empirical properties of the same-ZIP physician network.
Panel (a) shows the complementary cumulative distribution function of degree, revealing a
heavy-tailed structure with a small number of highly connected prescribers.
Panel (b) shows substantial persistence in network structure over time.
Panel (c) highlights pronounced geographic heterogeneity in opioid prescribing intensity
across counties, indicating meaningful variation in the network environment.}
\label{fig:overview}
\end{figure}

Figure~\ref{fig:overview} characterizes the network's structural
properties, temporal stability, and geographic heterogeneity. Panel (a)
shows a right-skewed degree distribution that was remarkably stable over
time. Panel (b) shows substantial persistence in network structure. Panel
(c) maps pronounced geographic heterogeneity in prescribing intensity
across states.

\subsection*{Evidence on Hypothesis 1: Peer Exposure and Prescribing Behavior}

\begin{figure}[H]
\centering
\includegraphics[width=0.85\textwidth]{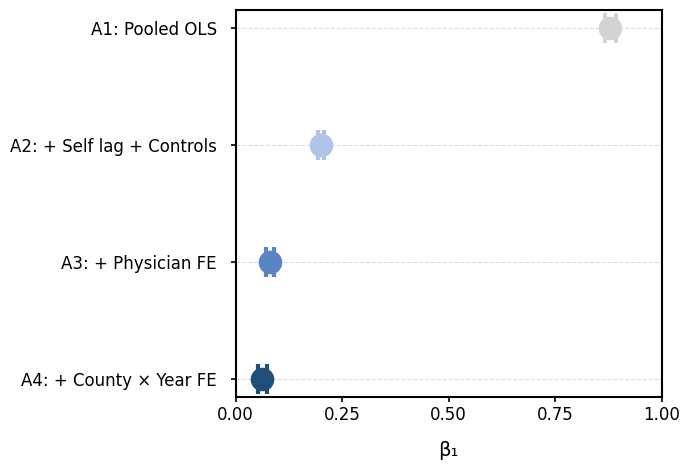}
\caption{Estimates of the relationship between physicians' prescribing and the average prescribing
of their peers across regression specifications. The figure reports estimates of the coefficient
$\beta_1$ from regressions of physician-level opioid prescribing on the average prescribing of
connected peers. A1: pooled OLS, no fixed effects. A2: adds own lagged prescribing and controls.
A3: adds physician fixed effects. A4: adds county-by-year fixed effects.
Horizontal bars denote 95\% confidence intervals. Standard errors clustered at the county level.}
\label{fig:peer_effect_regression}
\end{figure}

Figure~\ref{fig:peer_effect_regression} shows estimates from a sequence of
specifications that progressively address potential confounding. The pooled
OLS estimate indicated a strong positive association. After including
physician fixed effects and county-by-year fixed effects, the estimated
association remained positive, economically meaningful, and precisely
estimated.

Table~\ref{tab:peer_effect_table} reports the corresponding estimates. In
the preferred specification (A4), the estimated coefficient of peer
prescribing is 0.0631 ($SE = 0.0050$), interpretable as an elasticity: a
1\% increase in peer prescribing is associated with a 0.063\% increase in
individual prescribing in the subsequent period. Moving from the 10th to
the 90th percentile of the peer prescribing distribution yields an increase
of approximately 0.9\% in a physician's own prescribing.

\begin{table}[H]
\centering
\caption{Estimates of the relationship between physicians' prescribing and the average prescribing of
their peers across regression specifications.
Column A1 presents pooled OLS estimates. A2 adds the physician's own lagged prescribing and controls.
A3 further includes physician fixed effects. A4 includes county-by-year fixed effects.
All regressions are estimated at the physician-year level; standard errors clustered at the county level.}
\label{tab:peer_effect_table}
\begin{tabular}{lcccc}
\toprule
 & (A1) & (A2) & (A3) & (A4) \\
\midrule
Peer prescribing ($\bar{p}_{N(i),t}$)
  & 0.8801 & 0.2005 & 0.0812 & 0.0631 \\
SE
  & (0.0061) & (0.0035) & (0.0045) & (0.0050) \\
\midrule
Own prescribing, $p_{i,t}$
  &          & 0.7732 & 0.2253 & 0.2220 \\
SE
  &          & (0.0029) & (0.0040) & (0.0039) \\
\midrule
$\log(1+\text{patient volume}_{i,t})$
  &          & $-$0.0008 & $-$0.0004 & $-$0.0000 \\
SE
  &          & (0.0001) & (0.0003) & (0.0002) \\
\midrule
Physician FE        & No  & No  & Yes & Yes \\
County $\times$ Year FE & No & No & No  & Yes \\
Controls            & No  & Yes & Yes & Yes \\
\midrule
Observations & \multicolumn{4}{c}{2,005,267} \\
\bottomrule
\end{tabular}
\end{table}

\subsection*{Evidence for Hypothesis 2: Network Position and Differential Responses to Policy}

\begin{figure}[H]
\centering
\includegraphics[width=0.80\textwidth]{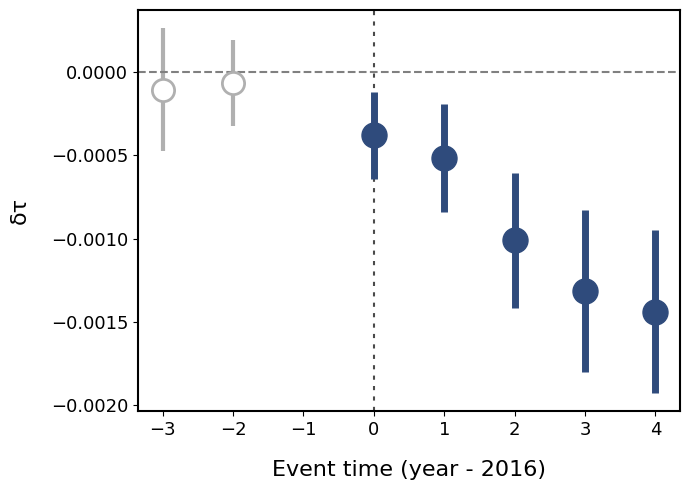}
\caption{Event study estimates of differential prescribing responses by network centrality around
the 2016 CDC opioid prescribing guideline. Coefficients $\delta_\tau$ are from regressions of
physician-level prescribing on event-time indicators interacted with a standardized (z-score)
measure of 2015 network centrality, with 2015 as the omitted baseline year. The vertical dashed
line indicates the timing of the guideline release. Pre-treatment coefficients for 2013 and 2014
are jointly indistinguishable from zero (F-test, $p = 0.85$). Error bars denote 95\% confidence
intervals. All regressions include physician and year fixed effects.}
\label{fig:event_study_regression}
\end{figure}

Figure~\ref{fig:event_study_regression} presents event study estimates of
prescribing responses by physician centrality. Pre-treatment coefficients
for 2013 and 2014 were small and statistically indistinguishable from zero
(F-test, $p = 0.85$), supporting the parallel trends assumption. Following
the introduction of the guideline, physicians with higher centrality
exhibited larger and increasingly pronounced reductions in prescribing
intensity.

Table~\ref{tab:event_study_table} reports the corresponding estimates.
Coefficients became negative and grew in magnitude from $-0.0004$ in 2016
to $-0.0014$ in 2020.

\begin{table}[H]
\centering
\caption{Event study estimates of differential prescribing responses by network centrality
around the 2016 CDC opioid prescribing guideline.
Coefficients represent the differential response associated with an increase of one SD in
network centrality. Pre-treatment estimates (2013 and 2014) are jointly indistinguishable
from zero ($p = 0.85$). All regressions are at the physician-year level; standard errors
clustered at the county level.}
\label{tab:event_study_table}
\begin{tabular}{lccc}
\toprule
Year & $\beta$ & SE & 95\% CI \\
\midrule
2013 & $-$0.0001 & (0.0002) & [$-$0.0005,\;0.0003] \\
2014 & $-$0.0001 & (0.0001) & [$-$0.0003,\;0.0002] \\
2016 & $-$0.0004 & (0.0001) & [$-$0.0006,\;$-$0.0001] \\
2017 & $-$0.0005 & (0.0002) & [$-$0.0008,\;$-$0.0002] \\
2018 & $-$0.0010 & (0.0002) & [$-$0.0014,\;$-$0.0006] \\
2019 & $-$0.0013 & (0.0002) & [$-$0.0018,\;$-$0.0008] \\
2020 & $-$0.0014 & (0.0003) & [$-$0.0019,\;$-$0.0009] \\
\midrule
Observations & \multicolumn{3}{c}{2,171,034} \\
\bottomrule
\end{tabular}
\end{table}

\begin{figure}[H]
\centering
\includegraphics[width=0.80\textwidth]{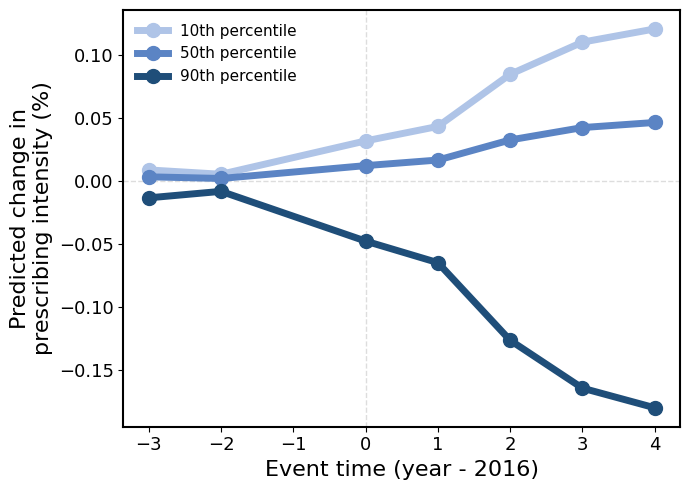}
\caption{Dynamic prescribing responses by baseline network centrality. Event time estimates of
predicted percentage changes in opioid prescribing relative to the 2015 baseline, separately for
physicians at the 10th, 50th, and 90th percentiles of the baseline centrality distribution.
Event time is defined relative to the 2016 CDC guideline. High-centrality physicians (90th
percentile) exhibited substantially larger reductions in prescribing following the guideline.
There is no evidence of differential pre-trends across groups.}
\label{fig:percentile_comparison}
\end{figure}

Figure~\ref{fig:percentile_comparison} plots event time coefficients
separately for physicians at the 10th, 50th, and 90th percentiles of the
baseline centrality distribution. By 2018, the difference in prescribing
between the 90th and 10th percentile physicians reached 0.2109\% and
continued to increase to 0.2742\% in 2019 and 0.3002\% in 2020 (expressed
as predicted percentage changes, $\exp(\Delta\log)-1$). The gap between
the 50th and 10th percentiles remained substantially smaller at 0.0520\%,
0.0677\%, and 0.0741\% over the same period.

\subsection*{Evidence for Hypothesis 3: Peer Adjustment and the Propagation of Prescribing Changes}

\begin{figure}[H]
\centering
\includegraphics[width=0.80\textwidth]{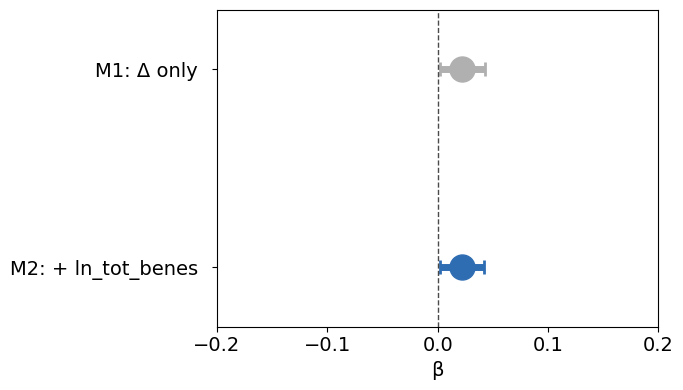}
\caption{Estimates of peer adjustment in prescribing behavior following the introduction of the
2016 CDC opioid prescribing guideline. Coefficients $\beta$ from first-difference regressions of
changes in physician-level prescribing intensity between 2016 and 2017 on corresponding changes
in network-weighted peer prescribing. M1 includes only changes in peer prescribing. M2
additionally controls for baseline patient volume. Points denote coefficient estimates; horizontal
bars represent 95\% confidence intervals. Standard errors clustered at the county level.}
\label{fig:mechanism_test}
\end{figure}

Figure~\ref{fig:mechanism_test} reports first-difference estimates.
Table~\ref{tab:mechanism_test_table} reports the corresponding estimates.
In specification M1, a one-unit change in network-weighted peer prescribing
is associated with a 0.0225 change in individual prescribing ($p = 0.030$).
The estimate remained stable in specification M2 after controlling for
baseline patient volume ($\beta = 0.0221$, $p = 0.032$), indicating the
relationship is not driven by differences in physician scale.

\begin{table}[H]
\centering
\caption{Peer adjustment and individual prescribing responses following the 2016 CDC opioid
prescribing guideline. M1 includes only changes in peer prescribing. M2 additionally controls
for baseline patient volume. Standard errors clustered at the county level.}
\label{tab:mechanism_test_table}
\begin{tabular}{lcccccc}
\toprule
 & $\beta$ & SE & $p$-value & 95\% CI & $N$ & Controls \\
\midrule
M1 & 0.0225 & (0.0103) & 0.030 & [0.0022,\;0.0427] & 246,481 & No \\
M2 & 0.0221 & (0.0103) & 0.032 & [0.0019,\;0.0424] & 246,481 & Yes \\
\bottomrule
\end{tabular}
\end{table}

\section{Discussion}

This study showed that opioid prescribing is shaped by professional
networks. Policy interventions appear to spread through these networks, producing responses that vary across physicians in ways that individual characteristics alone do not fully explain. All three results point toward a common mechanism:
network position shapes both routine prescribing behavior and the
transmission of policy-relevant behavioral change.

The stronger guideline responses observed among more central physicians are
consistent with a well-established finding that centrally positioned
professionals facilitate information transfer and serve as natural conduits
for behavioral change
\citep{tasselli2015social,cunningham2012health,soumerai1998effect}. Our findings extend aggregate evidence that the 2016 CDC guideline accelerated the decline in high-dose opioid prescribing \citep{bohnert2018opioid}. We show that these changes were concentrated among more central physicians and moved together with the prescribing of their local peers. By 2020, physicians at the 90th percentile of the centrality distribution exhibited prescribing reductions approximately 0.30 percentage
points larger than physicians at the 10th percentile, a gap that widened
steadily over the four years following the guideline's release.

The first-difference evidence is consistent with social learning models in
which agents update by incorporating the behavior of their neighbors
\citep{golub2010naive,chandrasekhar2020testing}. This pattern is also
connected to the clinical literature showing that peer comparison feedback
and local norm setting are among the most effective tools for changing
prescribing practices
\citep{zhang2020systematic,stepan2019development,wetzel2018interventions}.
The absorption of peer variation by county fixed effects in the
first-difference design underscores the general methodological principle
that identification strategies must align with the geographic level at
which peer interactions are defined \citep{marcus2021role,chiu2026causal}.

These findings have direct implications for opioid stewardship programs.
Network-based targeting strategies have been shown to generate substantial
cascading effects in public health contexts
\citep{alexander2022algorithms,kim2015social,valente2012network,valente2020diffusion}.
The theory of optimal network targeting predicts that interventions placed
on high-centrality nodes generate disproportionately large spillovers when
behaviors are strategically complementary \citep{galeotti2020targeting}.
Volume and centrality are related but distinct dimensions of physician
network position, and our results suggest that centrality may be a more
policy-relevant criterion for identifying physicians whose behavior changes
will propagate most broadly through the network
\citep{gouin2022identifying,iyengar2014social,hao2018empirical}. At the same time, lower prescribing is not an unambiguous good. Efforts to reduce prescribing that are too broad or poorly placed can leave patients with legitimate pain undertreated. A network-based strategy would therefore need to be paired with attention to appropriate care, and should not be judged by reductions in prescribing volume alone.

This study had several limitations. The physician network was built from geographic co-location within ZIP codes, which serves as a proxy for local professional interaction. This construction treats all physicians in the same ZIP code as connected. It does not capture actual communication, referral, or shared-patient relationships. Patient-sharing network data
\citep{donohue2018influence} would allow a more precise characterization of
interaction channels. The analysis relies on Medicare Part D data, which
primarily reflect prescribing for older and disabled populations; opioid
prescribing dynamics may differ in other settings \citep{ang2025expanded}.
Two issues limit causal interpretation. The reflection problem makes it hard to separate a physician's influence on peers from peers' influence on the physician \citep{manski1993identification}. Shared local conditions and the sorting of physicians into practice areas can also produce correlated behavior that is not peer influence \citep{goldsmith2013social}. Our physician and county-by-year fixed effects narrow these concerns for the peer-exposure results, and the event study leans on an external policy shock with no detectable pre-trends for the centrality results. Even so, we treat the peer estimates as associations rather than as fully identified causal effects. Our analysis also does not address the distributional consequences of network-based targeting. Concentrating attention on highly central physicians would shift where monitoring, education, and oversight fall. Where those physicians practice, and which patients they serve, carries equity implications that we do not measure here. Future work should ask whether such targeting narrows or widens existing disparities in access to pain care.

More broadly, this study contributes to a growing body of literature
showing that professional behavior is shaped by the structure of the
networks in which individuals are embedded
\citep{lee2025granular,weng2026rewiring,lee2026adaptive}. Future research
could extend this approach using richer network data, additional policy
shocks, or counterfactual simulations of alternative
intervention-targeting strategies.

Physicians do not adjust their prescribing in isolation. They respond to the behavior of peers and to their position within local professional networks. Policies that recognize this social structure, by engaging well connected physicians or using peer feedback, may reach further than those that treat physicians as independent actors. Whether they do so safely and fairly is a question for the next stage of this work.

\section*{Acknowledgments.}
%The authors thank participants in seminars at Lehigh University and
%National Taiwan University for valuable comments. H.-W.L.\ acknowledges
%support from the Lehigh University Office of the Vice Provost for Research.

\section*{Author Contributions.}
%Y.-N.W.\ and H.-W.L.\ designed the study, conducted the analysis, and
%wrote the manuscript.

\section*{Data Availability.}
All data are publicly available from the Centers for Medicare and Medicaid
Services (CMS) Medicare Part D Prescriber Public Use Files
(\url{https://data.cms.gov}). ZIP-to-county crosswalk files are available
from the HUD--USPS ZIP Code Crosswalk. Code for replication will be made
available upon publication.

\section*{Competing Interests.}
The authors declare no competing interests.

\bibliographystyle{unsrt}
%\vspace{-1.5cm}
\bibliography{ref}
\clearpage

\appendix
\renewcommand{\thefigure}{A.\arabic{figure}}
\setcounter{figure}{0}
\renewcommand{\thetable}{A.\arabic{table}}
\setcounter{table}{0}
\renewcommand{\theequation}{A.\arabic{equation}}
\setcounter{equation}{0}
\section{Supplementary Material}

\subsection*{Appendix: Alternative outcome measure (prescribing propensity)}

To further assess the robustness of the evidence for Prediction 1, we re-estimate the relationship between peer prescribing and individual prescribing using an alternative outcome measure: opioid prescribing propensity, defined as the share of opioid prescriptions among a physician's total prescriptions.

\begin{figure}[H]
   \centering
   \includegraphics[width=0.85\textwidth]{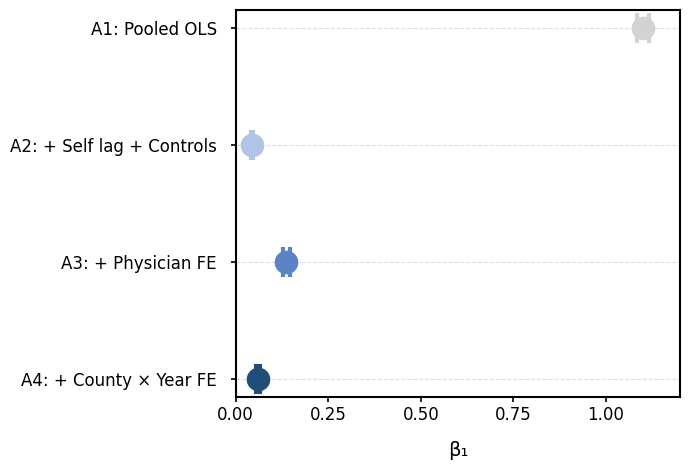}
   \caption{Estimates of the relationship between peer prescribing and individual prescribing using prescribing propensity as the outcome. The figure reports estimates of the coefficient ($\beta_1$) from regressions of physician-level opioid prescribing propensity on the average prescribing propensity of connected peers. Prescribing propensity is defined as the share of opioid prescriptions among a physician's total prescriptions. Specifications correspond to those in the main analysis: A1 (pooled OLS), A2 (adding own lagged prescribing propensity and controls), A3 (adding physician fixed effects), and A4 (adding county-by-year fixed effects). Points denote coefficient estimates, and horizontal bars represent 95\% confidence intervals. All regressions are estimated at the physician-year level, with standard errors clustered at the county level.}
   \label{fig:peer_effect_propensity}
\end{figure}

Figure \ref{fig:peer_effect_propensity} presents the corresponding estimates across specifications. The results closely mirror those obtained using prescribing intensity. The pooled OLS estimate indicates a strong positive association between peer prescribing propensity and individual prescribing propensity. As additional controls and fixed effects are introduced, the estimated coefficient declines in magnitude but remains positive and precisely estimated across all specifications.

The persistence of this relationship under an alternative outcome definition suggests that the main results are not driven solely by differences in overall prescribing volume. Instead, physicians appear to adjust the composition of their prescribing behavior in response to their peers, increasing or decreasing the relative share of opioid prescriptions within their overall practice.

Taken together, these findings provide additional evidence consistent with Prediction 1: physicians' prescribing behavior responds to the behavior of their peers. Moreover, the results indicate that such responses operate not only through changes in prescribing intensity but also through adjustments at the extensive margin of prescribing decisions.

\subsection*{Appendix: Alternative peer definition (county-level aggregation)}

To further assess the robustness of the evidence for Prediction 1, we construct an alternative measure of peer exposure based on geographic co-location at the county level. In this specification, a physician's peer prescribing is defined as the leave-one-out average prescribing of other physicians practicing in the same county-year. Unlike the baseline ZIP-code network, this approach does not rely on pairwise network links, but instead captures broader geographic exposure.

\begin{figure}[H]
   \centering
   \includegraphics[width=0.85\textwidth]{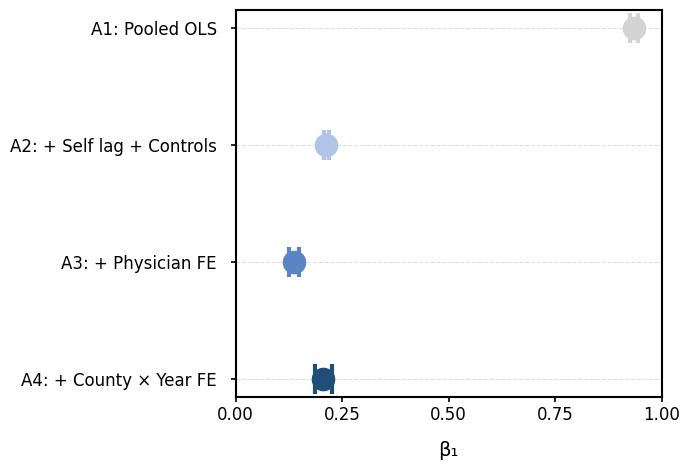}
   \caption{Estimates of the relationship between peer prescribing and individual prescribing using a county-level geographic peer definition. The figure reports estimates of the coefficient ($\beta_1$) from regressions of physician-level prescribing on the leave-one-out average prescribing of other physicians practicing in the same county-year. Unlike the baseline ZIP-code network, this measure captures peer exposure through geographic aggregation rather than pairwise network links. Specifications correspond to those in the main analysis: A1 (pooled OLS), A2 (adding own lagged prescribing and controls), A3 (adding physician fixed effects), and A4 (adding county-by-year fixed effects). Points denote coefficient estimates, and horizontal bars represent 95\% confidence intervals. All regressions are estimated at the physician-year level, with standard errors clustered at the county level.}
   \label{fig:same_county_network}
\end{figure}

Figure \ref{fig:same_county_network} reports the corresponding estimates across specifications. The results remain qualitatively similar to the baseline findings. The pooled OLS estimate is again large and positive, and although the magnitude declines with the inclusion of controls and fixed effects, the estimated relationship remains positive and statistically significant across all specifications.

Taken together, these findings provide additional evidence consistent with Prediction 1: physicians' prescribing behavior is systematically related to the prescribing behavior of nearby peers. The persistence of the relationship under a broader geographic definition of peer exposure suggests that the main results are not driven solely by the specific structure of the ZIP-code network. At the same time, the attenuation of the estimates relative to the baseline specification is consistent with the idea that more precisely measured network links capture stronger peer interactions than coarse geographic aggregation.

\subsection*{Appendix: Heterogeneity in the relationship between peer and individual prescribing}

As a further assessment of Prediction 1, we examine whether the relationship between peer prescribing and individual prescribing varies systematically across physician characteristics, including specialty, practice size, and geographic location. Table~\ref{tab:A2_heterogeneity} reports the corresponding estimates.

Panel~A considers heterogeneity by specialty. Physicians are grouped into primary care/internal medicine (baseline), pain-related specialties, and surgical/acute specialties. While the interaction coefficients vary modestly across groups, the implied total effects remain similar in magnitude across all specialties. This suggests that the relationship between peer prescribing and individual prescribing is not concentrated in specific clinical domains but instead operates broadly across specialties.

Panel~B examines heterogeneity by practice size, where physicians are divided into small and large practices based on the median number of beneficiaries. The interaction between peer prescribing and practice size is effectively zero across all specifications, indicating that the strength of the relationship does not vary meaningfully with the scale of practice.

Panel~C explores geographic heterogeneity by classifying physicians as urban or rural based on RUCA codes, where urban areas correspond to RUCA codes 1--3 and rural areas to codes 4 and above. The interaction term for urban physicians is close to zero and not statistically significant, suggesting that the relationship between peer prescribing and individual prescribing is comparable across geographic contexts.

Overall, the results provide little evidence of substantial heterogeneity across physician characteristics. Instead, the findings are consistent with Prediction 1 in indicating that the association between peer prescribing and individual prescribing is a pervasive feature of physician behavior rather than one limited to particular subgroups.

\begin{table}[H]
\centering
\small
\begin{tabularx}{\textwidth}{X cccc}
\toprule
 & (A1) & (A2) & (A3) & (A4) \\
\midrule
\midrule

\multicolumn{5}{l}{\textit{Panel A: Specialty}} \\
\addlinespace[2pt]

Peer effect (Primary/Medicine) 
& 0.87 (0.01) & 0.20 (0.00) & 0.08 (0.00) & 0.08 (0.00) \\

$\times$ Pain-related 
& 0.06 (0.00) & 0.01 (0.00) & -0.01 (0.00) & -0.01 (0.00) \\

$\times$ Surgical/Acute 
& -0.02 (0.00) & -0.01 (0.00) & 0.00 (0.00) & -0.00 (0.00) \\

Total effect (Pain-related) 
& 0.93 (0.01) & 0.21 (0.00) & 0.07 (0.01) & 0.07 (0.01) \\

Total effect (Surgical/Acute) 
& 0.85 (0.01) & 0.19 (0.00) & 0.08 (0.00) & 0.08 (0.00) \\

\addlinespace[6pt]
\multicolumn{5}{l}{\textit{Panel B: Practice Size}} \\
\addlinespace[2pt]

Peer effect (Small practice) 
& 0.88 (0.01) & 0.20 (0.00) & 0.08 (0.00) & 0.08 (0.00) \\

$\times$ Large practice
& 0.00 (0.00) & 0.00 (0.00) & 0.00 (0.00) & 0.00 (0.00) \\

Total effect (Large practice) 
& 0.88 (0.01) & 0.20 (0.00) & 0.08 (0.00) & 0.08 (0.00) \\

\addlinespace[6pt]
\multicolumn{5}{l}{\textit{Panel C: Urban vs Rural}} \\
\addlinespace[2pt]

Peer effect 
& 0.87 (0.01) & 0.20 (0.00) & 0.08 (0.00) & 0.08 (0.00) \\

$\times$ Urban 
& 0.01 (0.00) & 0.00 (0.00) & 0.00 (0.00) & 0.00 (0.00) \\

Total effect (Urban) 
& 0.88 (0.01) & 0.20 (0.00) & 0.08 (0.00) & 0.08 (0.00) \\

\bottomrule
\end{tabularx}
\caption{Heterogeneity in the relationship between peer prescribing and individual prescribing across physician characteristics. The table reports estimates from regressions that allow the association between peer prescribing and individual prescribing to vary by specialty, practice size, and geographic location. Panel~A groups physicians by specialty (primary care/internal medicine, pain-related, surgical/acute), Panel~B by practice size (split at the median number of beneficiaries), and Panel~C by location (urban vs.\ rural, where urban corresponds to RUCA codes 1--3 and rural to codes 4 and above). Reported coefficients include baseline estimates, interaction terms, and implied total effects for each subgroup. Standard errors are reported in parentheses.}
\label{tab:A2_heterogeneity}
\end{table}

\subsection*{Appendix: Event study with pre-determined network centrality}

\begin{figure}[H]
   \centering
   \includegraphics[width=1.0\textwidth]{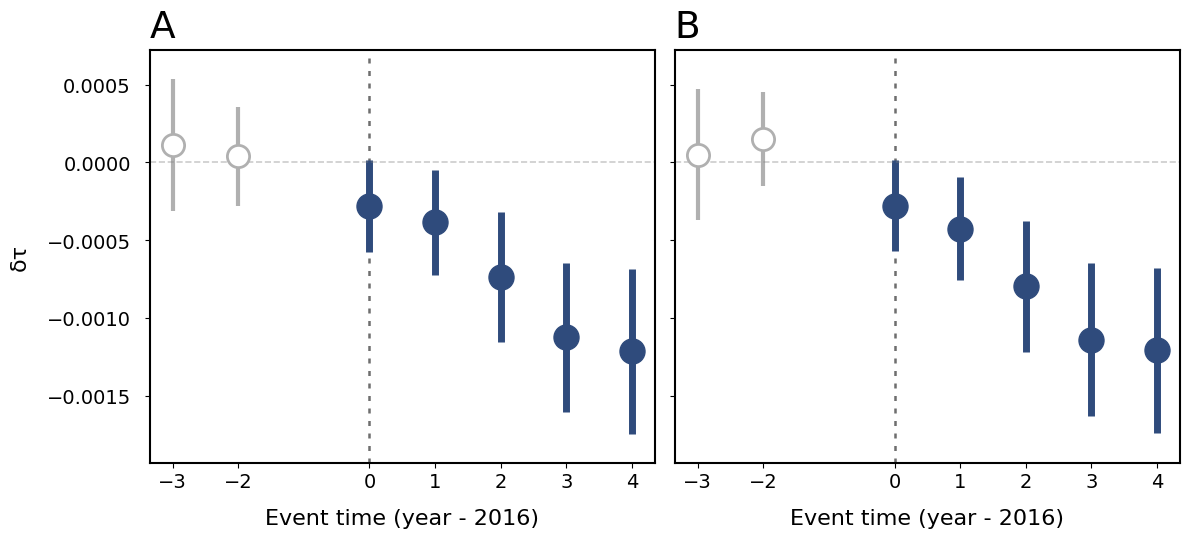}
   \caption{Event study estimates using pre-determined measures of network centrality. Panel (a) uses physician centrality measured in 2013, and Panel (b) uses centrality measured in 2014, both prior to the introduction of the 2016 CDC opioid prescribing guideline. The figure plots coefficients ($\delta_\tau$) from regressions of physician-level prescribing intensity on event-time indicators interacted with physician centrality. The vertical dashed line indicates the timing of the guideline. Pre-treatment coefficients are small and do not exhibit systematic trends in either panel. Following the policy change, coefficients become increasingly negative, indicating larger reductions in prescribing intensity among physicians with higher pre-policy centrality. The similarity of estimates across panels suggests that the results are not sensitive to the timing of centrality measurement. Error bars denote 95\% confidence intervals. All regressions include physician and year fixed effects, with standard errors clustered at the county level.}
   \label{fig:event_study_2013_2014}
\end{figure}

As a further assessment of Prediction 2, we re-estimate the event study using pre-determined measures of physician network centrality. Figure \ref{fig:event_study_2013_2014} presents results using centrality measured prior to the policy change, based on 2013 (Panel (a)) and 2014 (Panel (b)) values. Centrality is standardized (z-score) within each baseline year, without log transformation.

Across both specifications, the pre-treatment coefficients are small in magnitude and do not display systematic trends, consistent with the parallel trends assumption (F-test p = 0.83 for 2013 baseline and 0.39 for 2014 baseline). Following the introduction of the 2016 CDC opioid prescribing guideline, the coefficients become increasingly negative, indicating that physicians with higher pre-policy centrality experience larger reductions in prescribing intensity over time.

Importantly, the magnitude and dynamic pattern of the estimates are highly similar across the two panels. This consistency indicates that the results are not driven by the specific timing of centrality measurement. Instead, the findings support the interpretation that pre-existing differences in network position—measured prior to the policy shock—are systematically associated with heterogeneous responses to the guideline, consistent with Prediction 2.

\subsection*{Appendix: Event study using alternative centrality measures}

As a further assessment of Prediction 2, we re-estimate the event study using an alternative measure of physician network centrality. Instead of using weighted centrality based on the baseline ZIP-code network weights, we construct degree centrality, defined as the number of connections in the physician network in 2015. Centrality is standardized (z-score), without log transformation.

\begin{figure}[H]
   \centering
   \includegraphics[width=0.8\textwidth]{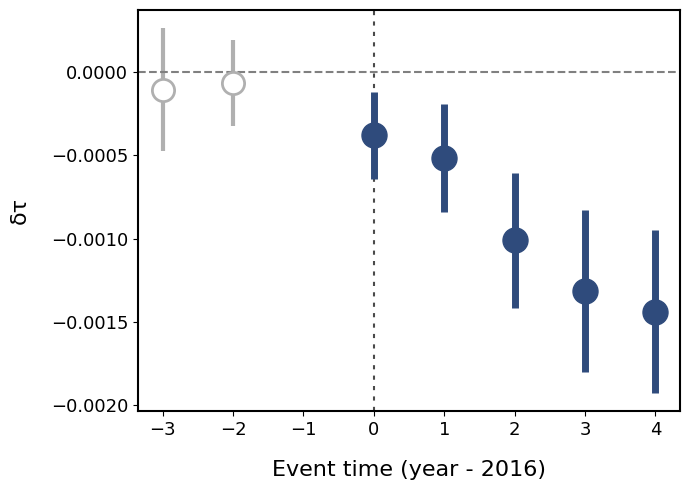}
   \caption{Event study estimates using degree-based network centrality. The figure plots coefficients ($\delta_\tau$) from regressions of physician-level prescribing intensity on event-time indicators interacted with physician degree centrality, defined as the number of connections in the network in 2015. The vertical dashed line indicates the timing of the 2016 CDC opioid prescribing guideline. Pre-treatment coefficients are close to zero and do not exhibit systematic trends. In the post-treatment period, coefficients become increasingly negative in the early years, reach their largest magnitude around 2018, and then partially attenuate. This dynamic pattern is consistent with a diffusion-type process in which more central physicians adjust earlier, followed by gradual convergence in prescribing behavior. Error bars denote 95\% confidence intervals. All regressions include physician and year fixed effects, with standard errors clustered at the county level.}
   \label{fig:event_study_degree}
\end{figure}

Figure \ref{fig:event_study_degree} presents the corresponding estimates. The results are qualitatively similar to those obtained using the baseline centrality measure. Pre-treatment coefficients remain close to zero and do not display systematic trends, providing no evidence of differential pre-trends (F-test p = 0.85).

In the post-treatment period, physicians with higher degree centrality experience larger reductions in prescribing intensity. The effect strengthens in the initial years following the 2016 guideline, reaches a peak around 2018, and subsequently attenuates in later periods. This non-monotonic pattern is consistent with a diffusion process in which more central physicians respond earlier, followed by gradual convergence in prescribing behavior.

Although the estimates are somewhat less precise than in the baseline specification, the overall dynamic pattern is preserved. These findings indicate that the results are not sensitive to the specific choice of centrality measure. Both weighted and unweighted measures of network position capture systematic differences in physicians’ responses to the policy change, consistent with Prediction 2.

\subsection*{Appendix: Event study using log-transformed centrality}

As a further assessment of Prediction 2, we re-estimate the event study using a log-transformed measure of physician network centrality. Specifically, we apply a $\log(1 + \text{centrality})$ transformation and standardize the resulting measure (z-score), based on the 2015 physician network.

\begin{figure}[H]
   \centering
   \includegraphics[width=0.8\textwidth]{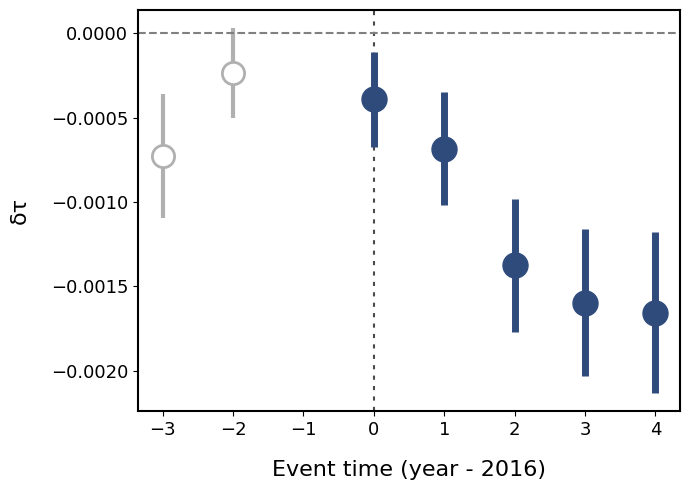}
   \caption{Event study estimates using log-transformed network centrality. Centrality is defined as $\log(1 + \text{centrality})$ and standardized (z-score) based on the 2015 network. The figure plots coefficients ($\delta_\tau$) from regressions of physician-level prescribing intensity on event-time indicators interacted with centrality, with 2015 as the omitted baseline year. The vertical dashed line indicates the timing of the 2016 CDC opioid prescribing guideline. Pre-treatment coefficients exhibit systematic deviations from zero, and a formal test rejects the null of parallel trends ($p = 0.0002$), indicating that this specification does not satisfy the identifying assumption. While post-treatment coefficients display a broadly similar pattern to the baseline specification, this violation of parallel trends motivates our use of the non-transformed centrality measure in the main analysis. Error bars denote 95\% confidence intervals. All regressions include physician and year fixed effects, with standard errors clustered at the county level.}
   \label{fig:event_study_zscore_log}
\end{figure}

Figure \ref{fig:event_study_zscore_log} presents the corresponding estimates. While the post-treatment coefficients exhibit a broadly similar dynamic pattern to the baseline specification, the pre-treatment coefficients display systematic deviations from zero. A formal test of parallel trends rejects the null of equal pre-treatment trends across centrality levels ($p = 0.0002$).

These results indicate that the log-transformed centrality specification does not satisfy the parallel trends assumption required for credible identification in the event-study framework. For this reason, we do not adopt this specification in the main analysis. Instead, we rely on the standardized (z-score) centrality measure without log transformation, which satisfies the parallel trends condition and provides a more credible basis for identifying heterogeneous responses, consistent with Prediction 2.

\subsection*{Appendix: First-difference specification with county fixed effects}

\begin{figure}[H]
   \centering
   \includegraphics[width=0.9\textwidth]{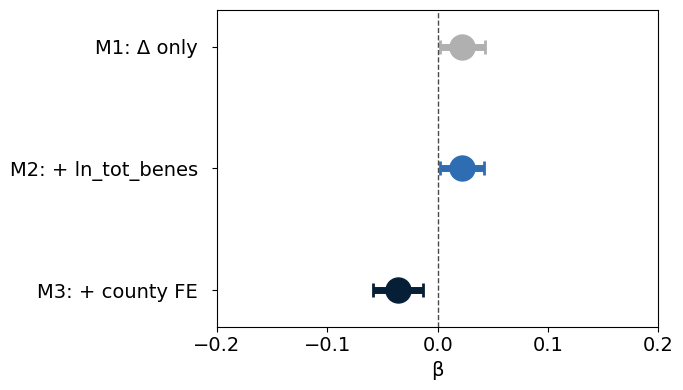}
   \caption{
    First-difference estimates of the relationship between changes in network-weighted peer prescribing and individual prescribing, with and without county fixed effects. 
    The figure reports coefficient estimates ($\beta$) from regressions of changes in physician-level prescribing intensity between 2016 and 2017 on corresponding changes in network-weighted peer prescribing constructed from the same-ZIP physician network. M1 includes only changes in peer prescribing, M2 additionally controls for baseline patient volume, and M3 further includes county fixed effects. While M1 and M2 yield positive and statistically significant estimates, the inclusion of county fixed effects in M3 substantially alters the coefficient, which becomes negative and statistically significant ($\beta = -0.0357$). This reflects the absorption of shared within-county variation in network-based peer exposure by county fixed effects, which reduces the identifying variation used to estimate peer effects. Points denote coefficient estimates, and horizontal bars represent 95\% confidence intervals. Standard errors are clustered at the county level.
    }
\label{fig:mechanism_county_fe}
   \label{fig:mechanism_test_county}
\end{figure}

As a further assessment of Prediction 3, we augment the first-difference specification by including county fixed effects. Figure \ref{fig:mechanism_test_county} reports the corresponding estimates alongside the baseline specifications.

Consistent with the main results, specifications without county fixed effects (M1 and M2) yield positive and statistically significant estimates, indicating that changes in peer prescribing are positively associated with changes in individual prescribing. In particular, the estimated coefficient is 0.0225 in M1 and 0.0221 in M2, both statistically significant at conventional levels.

However, once county fixed effects are introduced (M3), the estimated coefficient becomes negative and statistically significant (-0.0357). This pattern should not be interpreted as evidence of a reversal in peer effects, but rather reflects the reduction in identifying variation induced by the inclusion of geographic fixed effects.

Specifically, the peer prescribing measure is constructed from the same-ZIP physician network, using weighted exposure to connected peers. Because ZIP codes are nested within counties, physicians located in the same county are likely to share overlapping local peer networks and common geographic shocks. As a result, a substantial portion of the variation in network-weighted peer exposure is shared within counties.

The inclusion of county fixed effects absorbs much of this shared within-county variation in peer exposure, leaving only residual within-county differences for identification. These remaining differences are more limited and may be dominated by idiosyncratic variation, leading to estimates that are less stable and sensitive in sign and magnitude.

Therefore, the negative and statistically significant estimate in M3 should not be interpreted as evidence against peer effects. Rather, it reflects a limitation of the specification in which county fixed effects absorb key sources of variation in network-based peer exposure. For this reason, we do not adopt the county fixed effects specification in the main analysis, and instead rely on specifications that preserve more of the identifying variation in peer exposure to evaluate Prediction 3.

\subsection*{Appendix: Robustness of Predictions 1 and 3 to alternative edge-weighting schemes in network construction}

To assess the robustness of our results to alternative constructions of network ties, we re-estimate the main specifications underlying Predictions 1 and 3 using two alternative edge-weighting schemes. Specifically, we consider (i) an unweighted network in which all links receive equal weight, and (ii) an average-based weighting scheme that replaces the baseline minimum-based definition. In each case, peer exposure is recomputed based on the corresponding network and the main regressions are re-estimated.

Table \ref{tab:weight_robustness} summarizes the results for the main specifications. Across all edge-weighting definitions, the estimated peer effects remain positive and statistically significant in most specifications, with magnitudes closely comparable to those obtained under the baseline weighting. For Prediction 1 (A4), the coefficient on peer prescribing is $0.0631$ in the baseline specification, compared to $0.0587$ under the unweighted network and $0.0567$ under the average-based scheme. The differences relative to the baseline are small in magnitude, indicating that the estimated effect is stable across weighting choices.

A similar pattern emerges for Prediction 3 (M2). The estimated association between changes in peer prescribing and own prescribing remains positive across all weighting schemes, with coefficients of $0.0221$ (baseline), $0.0355$ (unweighted), and $0.0171$ (average-based). Although there is some variation in magnitude, all estimates are of the same sign and comparable order, and the differences relative to the baseline remain modest.

Overall, these results show that our main findings are not driven by the specific choice of edge-weighting function. Instead, they reflect a stable and consistently positive relationship between peer behavior and prescribing decisions that persists across alternative network constructions.

\begin{table}[H]
\centering
\begin{tabular}{lccc}
\toprule
 & Baseline & Unweighted & Avg-based \\
\midrule
\multicolumn{4}{l}{\textbf{Prediction 1 (A4)}} \\
\makecell[l]{Peer prescribing \\ (coefficient)}
& 0.0631 (0.0050) & 0.0587 (0.0047) & 0.0567 (0.0045) \\
\emph{Difference from baseline} 
& \textemdash & -0.0044 & -0.0064 \\
\\
\multicolumn{4}{l}{\textbf{Prediction 3 (M2)}} \\
\makecell[l]{$\Delta$ peer prescribing \\ (coefficient)}
& 0.0221 (0.0103) & 0.0355 (0.0165) & 0.0171 (0.0106) \\
\emph{Difference from baseline} 
& \textemdash & +0.0134 & -0.0050 \\
\bottomrule
\end{tabular}
\vspace{0.5em}
\footnotesize
\caption{Robustness to alternative edge-weighting schemes. This table summarizes the robustness of the main peer effect estimates for Predictions 1 and 3 using alternative definitions of network edge weights. The baseline specification defines edge weights as $w_{ij} = 0.1 \times \min(\text{Tot\_Benes}_i, \text{Tot\_Benes}_j)$. The ``Unweighted'' specification assigns equal weight to all edges, while the ``Avg-based'' specification uses an alternative averaging rule. For each weighting scheme, peer exposure variables are recomputed accordingly. The table reports results for the main specifications: Prediction 1 (A4) and Prediction 3 (M2). Across all weighting definitions, the estimated peer effects remain positive and statistically significant in most specifications, and their magnitudes are broadly comparable to those in the baseline specification. The reported differences relative to the baseline are small, indicating that the main results are robust to alternative constructions of network ties. Standard errors are reported in parentheses.}
\label{tab:weight_robustness}
\end{table}

\subsection*{Appendix: Placebo event-study estimates using 2015 as the policy year}

As an additional validation of our identification strategy, we conduct a placebo test by assigning the policy intervention to 2015 and re-estimating the event-study specification using only pre-policy data from 2013 to 2015. This exercise applies the same empirical framework as in the main analysis, with centrality measured based on 2014 (baseline year), but in a setting where no actual policy change occurred.

\begin{figure}[H]
   \centering
   \includegraphics[width=0.8\textwidth]{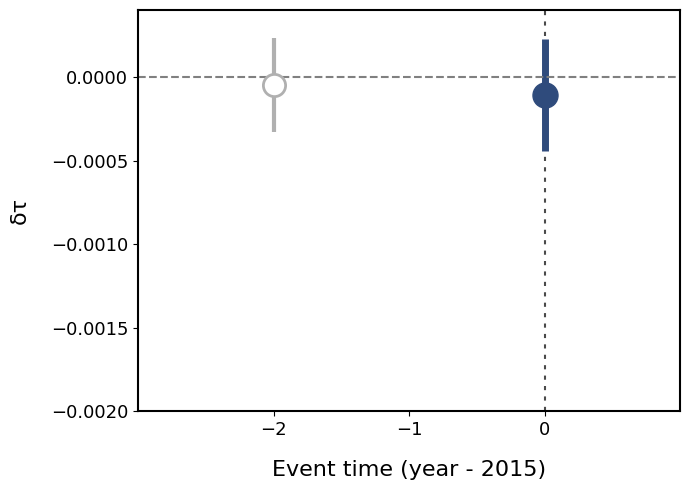}
   \caption{Placebo event study using 2015 as the policy year. This figure reports event-time estimates from a placebo version of the event-study specification, where the policy intervention is artificially assigned to 2015. The sample is restricted to the pre-policy period (2013–2015), with centrality measured based on 2014 (baseline year). The estimated coefficients are close to zero and statistically insignificant across all event times, with no systematic divergence between high- and low-centrality physicians. Consistent with this pattern, we fail to reject the null that all placebo event-time coefficients are jointly equal to zero (p = 0.73), providing no evidence of systematic differences across centrality groups. This contrasts with the main results following the actual 2016 CDC guideline and supports the interpretation that the observed heterogeneous response is driven by the policy intervention rather than by underlying trends.}
   \label{fig:2015_event_placebo}
\end{figure}

Figure~\ref{fig:2015_event_placebo} presents the corresponding estimates. In contrast to the main results, we find no evidence of systematic centrality-based divergence in prescribing behavior around the placebo policy date. The estimated coefficients are small in magnitude and statistically indistinguishable from zero, with no consistent differences between high- and low-centrality physicians over event time.

Importantly, we fail to reject the null that all placebo event-time coefficients are jointly equal to zero (p = 0.73). The absence of any emerging gap in the placebo period stands in sharp contrast to the pronounced and widening centrality gradient observed following the actual 2016 CDC guideline. This suggests that the heterogeneous response documented in Prediction 2 is not driven by pre-existing trends, gradual network evolution, or spurious time effects.

Taken together, these findings provide strong support for the interpretation that the observed centrality-based divergence is specifically induced by the policy shock, rather than reflecting a general tendency for more central physicians to adjust differently over time.

\subsection*{Appendix: Percentile-based differences across centrality percentiles for Prediction 2}

To quantify the magnitude of the heterogeneous responses across network positions, we translate the event-study estimates into differences between physicians at the 90th, 50th, and 10th percentiles of the baseline centrality distribution. Table~\ref{tab:percentile_interpretation} reports these differences for the post-policy years 2018--2020.

The results indicate substantial and economically meaningful divergence across centrality groups. By 2018, physicians at the 90th percentile of centrality exhibit a 0.0021 log-point larger decline in prescribing relative to those at the 10th percentile. This gap widens further to 0.0027 in 2019 and 0.0030 in 2020. Interpreted in percentage terms, this corresponds to approximately 0.21\%, 0.27\%, and 0.30\% larger declines in prescribing, respectively.

In contrast, differences between the 50th and 10th percentiles are much smaller, ranging from 0.0005 to 0.0007 log points, corresponding to approximately 0.05\% to 0.07\% differences in prescribing reductions. The gap between the 90th and 50th percentiles remains sizable, increasing from 0.0016 in 2018 to 0.0023 in 2020.

These results confirm that the centrality gradient is both large in magnitude and increasing over time. Highly central physicians respond substantially more to the policy shock, and the divergence relative to less connected physicians becomes progressively more pronounced in the years following the guideline.

\begin{table}[H]
\centering
\begin{tabular}{lcccc}
\toprule
Year & $\beta$ & p90--p10 & p50--p10 & p90--p50 \\
\midrule
2018 & -0.0010 & -0.0021 & -0.0005 & -0.0016 \\
2019 & -0.0013 & -0.0027 & -0.0007 & -0.0021 \\
2020 & -0.0014 & -0.0030 & -0.0007 & -0.0023 \\
\bottomrule
\end{tabular}
\caption{Percentile-based interpretation of centrality effects. This table reports differences in predicted prescribing changes across physicians at the 90th, 50th, and 10th percentiles of the baseline (2015) centrality distribution, based on the event-study estimates in Figure \ref{fig:percentile_comparison}. Differences are expressed in log points for post-policy years 2018--2020. Columns report pairwise differences between percentile groups: $p90 - p10$, $p50 - p10$, and $p90 - p50$. }
\label{tab:percentile_interpretation}
\end{table}

\subsection*{Appendix: Robustness of Prediction 2 to excluding highly central physicians}

As an additional robustness check for Prediction 2, we examine whether the estimated centrality gradient is driven by a small number of highly central physicians. To this end, we re-estimate the event-study specification after excluding the top 5\% of physicians in the centrality distribution.

Figure A.8 presents the corresponding estimates. The overall pattern closely mirrors the main results: following the 2016 policy intervention, prescribing declines more sharply among more central physicians, with a clear and persistent divergence across centrality groups.

Importantly, removing the most central physicians does not attenuate the estimated effects. The post-policy coefficients remain negative and of similar magnitude, and the dynamic pattern of divergence over event time is largely unchanged.

These results indicate that the centrality-based heterogeneity documented in Prediction 2 is not driven by a small set of extreme network nodes, but instead reflects a broader and more systematic relationship between network position and prescribing responses.

\begin{figure}[H]
   \centering
   \includegraphics[width=0.8\textwidth]{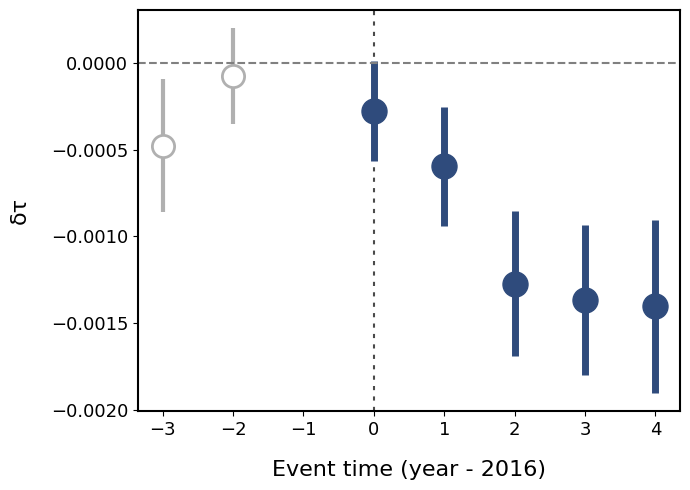}
   \caption{Excluding highly central physicians (top 5\%). This figure reports event-time estimates for Prediction 2 after removing the top 5\% of physicians in the centrality distribution. The specification follows the main event-study design, with centrality measured based on the baseline year. The estimated coefficients exhibit a similar dynamic pattern to the baseline results, with no systematic centrality-based divergence prior to the policy and a pronounced and persistent centrality gradient emerging following the 2016 policy intervention. The post-policy coefficients remain negative and comparable in magnitude, indicating that the main results are not driven by a small number of highly central physicians.}
   \label{fig:exclude_central_nodes}
\end{figure}

\end{document}